\documentstyle[pra,epsf,aps,twocolumn]{revtex}

\begin{document}
\sloppy
\def\bpi{\begin{picture}}
\def\epi{\end{picture}}
\def\comment#1{{}}
\author{H.~Kleinert\thanks{E-mail kleinert@physik.fu-berlin.de,~\newline
URL http://www.physik.fu-berlin.de/\~{}kleinert
}
                   \\
         Freie Universit\"at Berlin\\
          Institut f\"ur Theoretische Physik\\
          Arnimallee14, D-14195 Berlin
     }
\title{Fluctuation Pressure of Membrane between Walls}
\maketitle
\begin{abstract}
For a single membrane of stiffness $\kappa$
 fluctuating between two
planar walls of distance $d$,
 we calculate analytically the pressure law
\begin{equation}
p = \frac{\pi^2}{128}  \frac{k_B^2T^2}{ \kappa (d/2)^3}  .
\end{equation}
The prefactor ${\pi^2}/{128}   \sim 0.077115\dots$ is
in very good agreement with results from Monte Carlo
simulations $
0.079\pm 0.002$.
\end{abstract}
~\\
{\bf 1.} A stack of $n$ parallel, thermally fluctuating membranes exerts
upon the enclosing
planar walls a pressure which depends on the stiffness
$ \kappa$ and the temperature $T$ as follows
\begin{equation}
p =  \alpha_n \frac{2n}{n+1} \frac{k_B^2T^2}{ \kappa [d/(n+1)]^3}.
\label{1}\end{equation}
where $k_B$ is Boltzmann's constant
and $d$ the distance between the walls
(see Fig. 1).
\begin{figure}[tbhp]
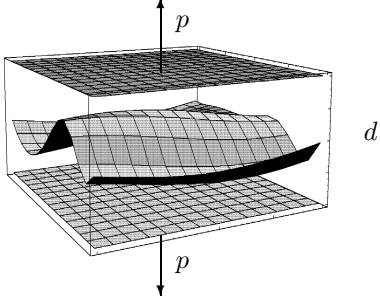

~\\
\input box.tps \\~~\\~~\\~\\  ~\\
\caption[]{Membrane fluctuating between walls
of distance $d$, exerting a pressure $p$.}
\label{@}\end{figure}

This law,
first deduced from
dimensional considerations by Helfrich \cite{1},
 is of fundamental importance in the statistical mechanics of
membranes just as the ideal gas law $pV= N\,k_BT$ in the
statistical mechanics of point particles.
We would therefore like to know the
 size of the prefactor,
the {\em stack constant\/}  $ \alpha _n$ as accurately as possible.
So far, its value
was determined only by
extensive
Monte Carlo
simulations as  being \cite{2,2b}
\begin{equation}
  \alpha_\infty = 0.101 \pm 0.002.
\label{@alMC}\end{equation}
For a single membrane, the following value was found \cite{3,2b}:
\begin{equation}
  \alpha _1 = 0.079 \pm 0.002.
\label{@MC}\end{equation}
So far, there exists no analytic theory to explain these values.

 The purpose  of this note is to fill this gap
for  the constant $ \alpha _1$,
by calculating analytically the pressure of a single membrane
between parallel walls.
The theoretical tool for this
has only recently become available:
A strong-coupling theory developed
originally
in
quantum mechanics \cite{Pfint},
was extended successfully to quantum field theories \cite{4},
where it has been used to obtain extremely accurate
values for  the critical exponents of O($n$)-symmetric scalar
fields with $\varphi^4$-interactions \cite{4}.

{\bf 2.} Strong-coupling theory gives direct access
to the large-$g$
behavior of divergent truncated power series expansions
of the type
\begin{equation}
 f_N (g)=  \Omega \left[ a_0 + \sum_{k=1}^{N} a_k\left(\frac{g}{ \Omega^q}\right)^k\right].
\label{EN1}\end{equation}
The
$g \rightarrow \infty$ -limit
of $f_N(g)$, to be denoted by
$f_N^*$,  is obtained by setting $ \Omega\equiv cg^ {1/g}$ and optimizing the
function
\begin{equation}
f_N(c)= g^{1/q}\tilde f_N (c) \equiv g^{1/q} \left(
  c a_0 b^N_0 + \sum_{k=1}^{N} a_k c ^{-qk} b^N_k\right)
\label{EN}\end{equation}
where
\begin{equation}
  b_k^N = \sum_{l=0}^{N-k} (-1)^l \left( (1-kq)/2\atop l \right)
\label{@bnk}\end{equation}
is the binomial expansion of $(1-1)^{(1-kq)/2}$ truncated after the
$(N-k)$th term.
Optimizing means extremizing $\tilde f_N(c)$ in $c$ or,
if an extremum does not
exist, extremizing the derivative $\tilde f'_N(c)$.

{\bf 3.} We apply this theory to a membrane between walls by proceeding
as follows. The partition function of the membrane is given by the functional
integral
\begin{equation}
 Z = \int{\cal  D}\,u(x) \exp\left\{ -\frac{ \kappa}{2k_BT} \int d^2x [
\partial^2   u(x)]^2\right\}\equiv e^{-Af/k_BT},
\label{@Z}\end{equation}
where $u(x)$ is a vertical displacement field of the membrane fluctuating
between horizontal walls at $u= -d/2$ and $d/2$.
The quantities $A$ and $f$ are the wall area
  and the free energy per unit area, respectively.
Such a restriction of a field is hard to treat analytically.

We therefore perform a transformation which maps the interval
$u\in (-d/2, d/2)$ to an infinite $\varphi$-axis,
\begin{eqnarray}
 u &=& \frac{d}{\pi}\arctan\frac{\pi \varphi}{d}=
\varphi\left(1-\frac{\pi^2\varphi^2}{3d^2}
+\frac{\pi^4\varphi^4}{5d^4}+
\dots~\right).
\label{uphi}\end{eqnarray}
and add to the fluctuation energy $E$
in the exponent of (\ref{@Z}) a potential energy
which keeps the membrane between $-d/2$ and $d/2$ (P\"oschl-Teller potential):
\begin{eqnarray}
 E^{\rm pot} &=&
 E_0^{\rm pot}+
   E^{\rm int}=
\frac{ \kappa }{2} \int d^2 x\,{m^4}{\phi^2(u(x))}\label{E}\\
&=&
\frac{ \kappa m^4}{2} \int d^2 x\left\{ u^2(x)+
\sum_{k=2}^\infty \varepsilon_ k \left[\pi\frac{ u(x) }{d}\right]^{2k}\right\} ,
\nonumber \end{eqnarray}
with expansion coefficients $\varepsilon_2,\varepsilon_3,\varepsilon_4,\dots~$:
\begin{eqnarray}
{{1\,{}}\over 3},\,{{17}\over 90},
  \,{31\over 315},\,{691\over 14175},\,{10922\over 467775},\,
~\dots~.
\label{@expcoef}\end{eqnarray}
\begin{figure}[tbhp]
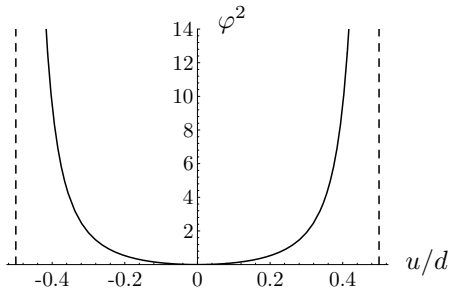

~~\\
\input pot.tps
\caption[]{Smooth Potential replacing box walls}
\label{@}\end{figure}%
The potential energy per area is plotted in Fig.~1. Its presence
destroys the simple scaling properties
of the partition function
(\ref{@Z}), which depends only on the dimensionless variable
 $\kappa d^2/k_BT$.
The new partition function $Z$ associated with
the modified energy  $E+E^{\rm pot}$
has an additional dependence on the dimensionless variable
$g=\pi^2/m^2d^2$. The original hard-wall system
is obtained
in the strong-coupling limit $g\rightarrow \infty$.

In  the opposite limit where $g$ goes to zero,
the
energy $E+E^{\rm pot}$ becomes harmonic,
\begin{equation}
 E_0 = \frac{\kappa}{2} \int d^2 x\{ [\partial^2u(x)]^2+m^4u^2(x)\},
\label{@}\end{equation}
leading to a partition function
\begin{equation}
Z_0 = e^{-\frac{1}{2}{\rm Tr }\log (\partial^4 + m^4) } = \mbox{const}
     \times e^{-\frac{A}{8} m^2} ,
\label{@12}\end{equation}
where $A$ is the area of the walls.

For a finite distance $d$, the
interaction energy $E^{\rm int}$
is treated perturbatively order by order in $g$,
expanding the exponential $e^{-E^{\rm int}/k_BT}$ in a power series,
and each power in a sum of all pair contractions.
These are pictured
by loop diagrams
whose lines represent
the correlation function
\begin{equation}
 \langle
u(x_1)
u(x_2)
\rangle = \frac{ \kappa}{k_BT} \int \frac{d^2k}{ (2 \pi)^2}  \frac{1}{k^4+m^4}
    e^{i{ k} (x_1 - x_2) }.
\label{CF}\end{equation}
The free energy density $f=-k_BTA^{-1}\log Z$ is obtained
from  all connected loop diagrams. For simplicity, we shall
use natural units with $ \kappa /k_BT=1$.

The lowest contribution to the free energy density
comes from the expectation value of the
 $u^4$-interaction  or the loop diagram
$3\,$
\hspace{-10pt}
\rule[-10pt]{0pt}{26pt}\unitlength.2mm
\begin{picture}(42,12)
\put(15,6){\circle{16}}
\put(31,6){\circle{16}}
\put(23,6){\circle*{4}}
$\!$\end{picture}
which is
of the order $1/d^2$:
\begin{equation}
\frac{ m^4}{2d^2}\langle u^4\rangle
=\frac{ m^4}{2d^2}3\langle u^2\rangle^2,
\label{@}\end{equation}
the line representing the pair expectation
\begin{equation}
\langle u ^2 \rangle =
 \int \frac{d^2k}{ (2 \pi)^2}  \frac{1}{k^4+m^4}
   = \frac{1}{8m^2}.
\label{@m1}\end{equation}
Together with the exponent in (\ref{@12}), we thus obtain
first-order free energy density
\begin{equation}
f_1 = \frac{m^2}{8} + \frac{1}{32} \frac{ \pi^2}{m^2d^2}.
\label{@f1m}\end{equation}
Continuing the perturbation expansion,
yields
an expansion of the
general form
\begin{eqnarray}
  f_N &=& m^2 \left[ \frac{1}{8} \!+\! \frac{1}{64}
\frac{ \pi^2}{m^2d^2}\!+\! \right.
	 \nonumber \\&&~~~~~~+\left.a_2 \left( \frac{ \pi^2}{m^2d^2}\right)^2 \!+\!
	 \dots\!+\!a_N \left( \frac{ \pi^2}{m^2d^2}\right)^N\right] ,
\label{@f1mm}\end{eqnarray}
where $a_2,\dots,a_N$ are dimensionless numbers.
By comparison
with
(\ref{EN1}) we identify $p=q=1,  ~\Omega=m^2$,~
$g =  \pi^2/d^2$.
 The function $ f_N(c)$ of Eq.~(\ref{EN})
describing the limiting large-$g$
 behavior
 is obtained by
 setting
$ \Omega \equiv c\pi^2/2d^2$, and reads
\begin{equation}
f_N(c) = \frac{\pi^2}{d^2}\left(
\frac{c}{4}b_0^N + \frac{1}{64} +
 \frac{a_2}{c}	 b_2 ^N
+\dots
 +\frac{a_N}{c^{N-1}}	 b_N ^N
\right) .
\label{fb0}\end{equation}
According to the above-described
strong-coupling theory, we must optimize
the  expression $\tilde f_N(c)$
in parentheses. Since the second term does not contain $c$,
we separate this term out, and write
\begin{equation}
\tilde f_N(c)\!=\!
\frac{1}{64}\!+\!
 \Delta {\tilde f_N}(c)\!\equiv \!
\frac{1}{64}\!+\!\left(\frac{c}{4}b_0^N\!+\!\frac{a_2}{c}b_2^N\!+\!\dots
\!+\!\frac{a_N}{c^{N-1}}	 b_N ^N
\right).
\label{@firstterm}\end{equation}
with only the remainder  $\Delta {\tilde f_N}(c)$
to be optimized.
Let $ \Delta \tilde f^*_N$ be ist optimal value.
If we know only $a_2$, we find the
approximation
$ \Delta f^*_2= \sqrt{{3a_2}/{16}} $.
Ignoring $ \Delta f^*_N$ for a moment,
the
first term in (\ref{@firstterm}) yields
the lowest estimate
for the free energy density of the original system
\begin{equation}
f_1^*=\frac{\pi^2}{64}\frac{1}{d^2},
\label{@f1m0}\end{equation}
implying a pressure law
\begin{equation}
p=-\frac{\partial f}{\partial d}= \frac{\pi^2}{32}\frac{1}{d^3}.
\label{@pre}\end{equation}
By comparison with
the general pressure law (\ref{1}), we identify the
prefactor as being
\begin{equation}
 \alpha _1=\frac{1}{2}\times \frac{\pi^2}{128}\approx\frac{1}{2}\times 0.077115.
\label{@alpha}\end{equation}
Without the prefactor factor $1/2$, this
would agree perfectly with the
Monte Carlo value (\ref{@MC}).
Thus we expect the
contribution of $ \Delta \tilde f_N^*$ for $N\rightarrow \infty$
to be equal or almost
equal to $1/64$.

The calculation of the  higher-order terms $a_2,\,a_3,\,\dots~$ is   tedious,
and will be presented in a separate detailed publication \cite{BK}.
In this note we shall circumvent it by exploiting
a close relationship
of the present problem with
a closely analogous exactly
solvable one, which may be treated in precisely the same way:
The euclidean version of a quantum-mechanical point particle in a one-dimensional box
 $u \in (-d/2,d/2)$.

{\bf 4.} The partition function of a particle in a box is
%
\begin{equation}
 Z = \int    {\cal D} u e^{-({ \kappa }/{2k_BT}) \int dx (\partial u)^2}
\equiv e^{-Af/k_BT}.
\label{@pp}\end{equation}
The quantum-mechanical ground state energy of this system is
exactly known: $(k_BT/ \kappa )\pi^2/2d^2$,
corresponding to a free energy density
\begin{equation}
 f = \frac{k_B^2T^2}{ \kappa }
\frac{\pi^2}{2d^2}.
\label{@}\end{equation}
The path integral
(\ref{@pp}) may now be treated as before, i.e.,
we transform $u$ to $\varphi$ via (\ref{uphi}), and
separate the field energy into  a Gaussian energy (in natural units)
\begin{equation}
 E_0 = \frac{ \kappa }{2} \int d x\{ [\partial^2u(x)]^2+ m^4u^2(x)\}
\label{@}\end{equation}
and
an interaction energy
which looks the same as (\ref{E}), except that the integration $\int d^2x$
runs now only over one dimension, $\int dx$.

The first-order contribution
to the free energy density
is now
(in natural units with $ \kappa /k_BT=1$)
\begin{equation}
\frac{  m^4}{2d^2}\langle u^4\rangle
=\frac{ m^4}{2d^2}3\langle u^2\rangle^2,
\label{@}\end{equation}
with the
pair expectation
\begin{equation}
\langle u ^2 \rangle = \int \frac{dq}{2 \pi}
  \frac{1}{q^2 +  m^4} = \frac{1}{2 m^2 },
\label{@}\end{equation}
leading to a first-order free energy density
\begin{equation}
f_1 =   \frac{m^2 }{2} + \frac{1}{2}\frac{  \pi^2}{  d^2}  ,
\label{W20}\end{equation}
and a
full
perturbation expansion of the form
\begin{equation}
f =  m^2 \left(\frac{1}{2} + \frac{1}{2}\frac{  \pi^2}{ m^2d^2}
 +a_2 \frac{ \pi^4}{ m^4 d^4} +\dots~ \right),
\label{W2}\end{equation}
From this we find the function $ f_N(c)$ defined in Eq.~(\ref{EN})
 governing the strong-coupling limit $d\rightarrow 0$ by setting
$ \Omega =m^2\equiv c\pi^2/2d^2$:
\begin{equation}
f_N(c) = \frac{\pi^2}{d^2} \tilde f_N(c),
\label{fb}\end{equation}
with
\begin{equation}
\tilde f_N(c) \!=\!\frac{1}{4}
\!+\! \Delta \tilde f_N(c)\!\equiv\!  \frac{1}{4}
\!+\!\left( \frac{c}{4}b_0 ^N \!+\! \frac{a_2}{c}	 b_2 ^N\!+\!\dots
 \!+\!\frac{a_N}{c^{N-1}}	 b_N ^N
\right) .
\label{fb}\end{equation}
Here the first term yields
the lowest approximation
\begin{equation}
   f_1 = \frac{1}{4}\frac{\pi^2}{d^2},
\label{@}\end{equation}
\unitlength.2mm
 which is {\em precisely\/} half
the exact result.
Thus we conclude that
the optimal value of the neglected expression
$ \Delta f_N(c)$ must be once more equal to $1/4$
in the limit $N\rightarrow \infty$.
In order to see how this happens,
we extend the Bender-Wu
recursion relation for the perturbation coefficients
of the anharmonic oscillator \cite{BW}.
It yields for the ground state energy an expansion%
{\footnotesize\begin{eqnarray}
&&\frac{1}{2}+\frac{3\pi^2}{4d^2} \varepsilon _4
-\frac{\pi^4}{8d^4}\left(21 \varepsilon^2 _4 -
15 \varepsilon _6\right)
\nonumber \\
&&\!+\frac{\pi^6}{16d^6}\left(
333
 \varepsilon _4^3 \!-\!
360 \varepsilon _4 \varepsilon _6
\!+\!105 \varepsilon _8\right)\nonumber \\
&&\!-\frac{\pi^8}{128d^8}\left(
30885
 \varepsilon _4^4
\!-\!44880 \varepsilon^2_4 \varepsilon _6
\!+\!6990 \varepsilon _6^2
\!+\!1512 \varepsilon _4 \varepsilon _8
\!+\!3780 \varepsilon _{10}
\right)\!+\!\dots\,.
\label{@}\nonumber \end{eqnarray}}%
Inserting the coefficients
(\ref{@expcoef}) we find
$a_2,a_4,a_6\dots~$:
\begin{equation}
      \frac{1}{16},\!-\frac{1}{256},
\,\frac{1}{2048},
\,\!-\frac{5}{65536},
\,\frac{7}{524288},
\,\!-\frac{21}{8388608},
\dots~,
\label{@}\end{equation}
whereas the odd coefficinets $a_3,a_5,a_7,\dots~$ vanish.
To account for this fact,
we  resum the series containing only the even terms
\begin{equation}
  \Delta \tilde f_{N} =\frac{ c }{4}
+\frac{a_2}c b^{N}_1
+\frac{a_4}{c^3} b^N_2\
+\frac{a_{2N}}{c^5} b^{N}_{N}
,
\label{@the func}\end{equation}%
taking the coefficients $b^N_i$ of Eq.~(\ref{@bnk})
with the parameters $p=1,\,q=2$.
 This yields the
functions
$ \Delta \tilde f_N(c)$ plotted in Fig.~3 and
listed in Table~1.
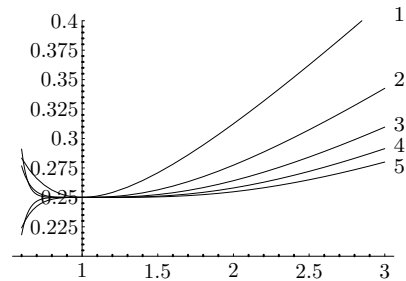
\begin{figure}[hb]
~\\    \unitlength.35mm
\input plsfall.tps
\caption[]{Plots of the functions  $ \Delta \tilde f_N(c)$
of Eq.~(\protect\ref{@the func}), all being optimal exactly at $c=1$
with $ \Delta \tilde f_N^*$=1/4.\\[-3mm] }
\label{@}\end{figure}
For all
 $ \Delta \tilde f_N(c)$,
optimization yields
a strong-coupling value
 $ \Delta \tilde f_N^*$
equal to $1/4$, thus raising the
initial value $1/4$ in  (\ref{fb}) to the correct final value $1/2$.

{\bf 5.}
To exploit this
property
of a particle in a box
for the system at hand, the
membrane between walls,
we make the following
crucial observation:
The Feynman integrals determining the first two terms
in the free energy densities
in Eq.~(\ref{@f1m}) for a membrane and in Eq.~(\ref{W20}) for a particle
 are related  to each other
by a simple transformation
of the integration variables.
The membrane integrals
\begin{eqnarray}
\int\!\!\!\frac{ d^2k}{(2\pi)^2}\log (k^4 + m^4) \!=\!\frac{m^2}{4},~~
\!\!\! \int\!\!\! \frac{d^2k}{ (2 \pi)^2}  \frac{1}{k^4+m^4}
   \!=\! \frac{1}{8m^2}
 \label{@memb}\end{eqnarray}
go over into those of the particle in the box
\begin{eqnarray}
\int \frac{dq}{2\pi}\log (q^2 +  m ^4) ={m^2 },~~~~
 \int \frac{dq}{ 2 \pi}  \frac{1}{q^2+ m^4}
   = \frac{1}{2 m^2 }
 \label{@part}\end{eqnarray}
by the transformation
$$k^2\rightarrow q,~~~~~\int\frac{d^2k}{(2\pi)^2}\rightarrow\frac{1}{4} \int _{-\infty}^\infty \frac{d q}{2\pi}.$$
Thus, if we multiply each loop integral
by a factor $1/4$,
we find immediately  the free energy density
$f_1$
of the membrane
in Eq.~(\ref{@f1m}) from that of the particle in the box in Eq.~(\ref{W20}).

\begin{figure}[tbh]
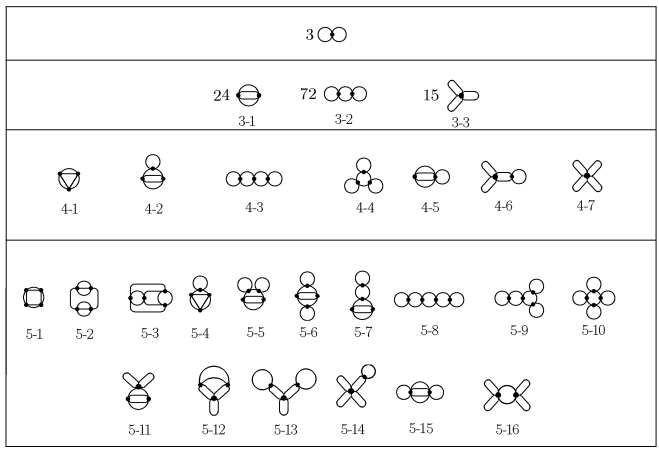

~\\
~\\
~\\
~\\
\input feynman.tps
\caption[Vacuum diagrams up to five loops
]{Vacuum diagrams up to five loops.
}
\label{diag}\end{figure}

But the analogy carries further: By differentiation
(\ref{@memb}) and (\ref{@part}) with respect to
$m^2$, we see that also all Feynman integrals
$ \int {d^2k}/ (2 \pi)^2{(k^4+m^4)^\nu}$ are related to the
$ \int {dq}/ (2 \pi)  {(q^2+m^4)^\nu}$
by the same factor $1/4$.
This property has the consequence
that {\em most\/} of the connected
loop diagrams
contributing to the perturbation expansion
of the free energy density,
shown in Fig.~\ref{diag} up to five loops,
are related by a factor
$(1/4)^L$, where $L$ is the number of loops.
In particular, all such diagrams
coincide which
are ususally summed in the Hartree-Fock approximation
(chain diagrams, daisy diagrams, etc.).
Only the topologicall more involved diagrams 3-1,~4-1,~4-2,\,4-5,~5-2,\,5-3,\,5-5,\,5-6,\,5-7,\,5.11,\,5-12,\,5-15
in   Fig.~\ref{diag} do not follow this pattern.
For a particle in a box, we can easily calculate the
associated Feynman integrals in $x$-space
as described in Chapter 3 of Ref.~\cite{Pfint},
and find that
they
contribute
less than 5\% to the sum of all diagrams at each loop level.
This implies that the corresponding results for
the membrane between walls will differ at most by this relative amount
from those for the particle in the box.
We therefore
conclude
that
since
the optimal value of
$ \Delta \tilde f_N(c)$ in Eq.~(\ref{fb}) doubles the initial value
for $N\rightarrow \infty$,
the  analogous function
for the membrane
between walls in Eq.~(\ref{@firstterm}) will double
approximately.
For the quantitative deviations
see
the forthcoming publication \cite{BK}.
A precise doubling of the result (\ref{@alpha})
leads to a very good agreement with the Monte Carlo number
(\ref{@alMC}).

{\bf 6.} The alert reader will have noted that
the field transformation (\ref{uphi})
is rather special.
We may, for instance, chose any mapping
\begin{eqnarray}
 u &=&  \frac{\varphi}{ \left[1 +  8\pi^2  \varphi^2/3d^2 +w_4  \varphi^4/d^4
+\dots
    +\left(2\varphi/d\right)^n\right]^{1/n} }
\nonumber \\&=&
\varphi-
\frac{2}{3}{\frac{\pi^2}{d^2}}\varphi^3
+\dots +{\cal O}(\varphi^5),
\label{uphi2}\end{eqnarray}
which has a doubled coefficient
of $\varphi^3$ with respect to the expansion
(\ref{uphi}). As a consequence, the functions $\tilde f_N(c)$ in
(\ref{@firstterm}) and (\ref{fb})
would have a doubled first term.
Since this would be the correct final value,
 the remaining functions
$ \Delta \tilde f_N(c) $  would have to converge to a vanishing optimal value
for $N\rightarrow \infty$ (in the particle case exactly,
in the membrane case approximately).
To reach this goal,
the coefficients $w_4,w_6,\dots~$ in (\ref{uphi2})
can be chosen rather arbitrarily, although there are
a few convenient ways for which the speed of convergence is
fast. A  preferred choice is one
in which all coefficients $a_2,a_3,a_3,\dots~ $
of the perturbation expansion vanishes for a particle in a box.
This and other possibilities will be studied separately \cite{5,tsch}.
~\\
~\\
Acknowledgement: \\
The author is grateful to
Prof.~A.~Chervyakov,
Prof.~B.~Hamprecht,
 Dr.~A.~Pelster, and M.~Bachmann for numerous
useful discussions.
\begin{table}[tbhp]
\caption[]{The functions $ \Delta \tilde f_N(c)$
of Eq.~(\protect\ref{@the func}), and their optimal values
 $ \Delta \tilde f_N^*$.}
\begin{tabular}{clc}
N&$ \Delta \tilde f_N(c)$& $ \Delta \tilde f_N^*$\\
\hline
1&$\frac{c}{4}+\frac{c^{-1}}{16c}$&$\frac{1}{4}$\\
2&$\frac{3c}{16}+\frac{3c^{-1}}{32}-\frac{c^{-3}}{256}$&$\frac{1}{4}  $\\
3&$\frac{5c}{32}+\frac{15c^{-1}}{128}-\frac{5c^{3}}{512}+\frac{c^{-5}}{2048}$&$\frac{1}{4}  $\\
4&$\frac{35c}{256}+\frac{35c^{-1}}{256}-\frac{35c^{3}}{3048}+\frac{7c^{-5}}{4096}-\frac{5}{65536}$&$\frac{1}{4}  $\\
5&$\frac{63}{512}+\frac{315c^{-1}}{2048}-\frac{105c^{3}}{4096}
+\frac{63c^{-5}}{16384}
-\frac{45c^{-7}}{131072}+\frac{7c^{-9}}{524288}$&$\frac{1}{4}  $
\end{tabular}
\label{@}\end{table}
\comment{
\begin{figure}[tbh]
\unitlength.15mm
\begin{center}
\footnotesize\begin{tabular}{|c||c|c|}
\hline
\hspace{-10pt}
\begin{tabular}{c}
\end{tabular}
\hspace{-10pt}
&
diagrams and multiplicities &$\!\!\!$number$\!\!\!$
\\
\hline
\hline
$g^1$&
$\!\!\!3$
\hspace{-10pt}
\rule[-10pt]{0pt}{26pt}
\bpi(42,12)
\put(13,3){\circle{16}}
\put(29,3){\circle{16}}
\put(21,3){\circle*{4}}
\epi&3
\\
\hline
$g^2$&
$\displaystyle\frac{1}{2!}\Bigg($
$\!\!\!24~$
\hspace{-10pt}
\rule[-14pt]{0pt}{34pt}
\bpi(34,12)
\put(17,3){\circle{24}}
\put(17,3){\oval(24,8)}
\put(5,3){\circle*{4}}
\put(29,3){\circle*{4}}
\epi
\hspace{10pt}
$\!\!\!72~$
\hspace{-10pt}
\rule[-10pt]{0pt}{26pt}
\bpi(58,12)
\put(13,3){\circle{16}}
\put(29,3){\circle{16}}
\put(45,3){\circle{16}}
\put(21,3){\circle*{4}}
\put(37,3){\circle*{4}}
\epi
$\Bigg)$&96
\\
\hline
$g^3$&
$\displaystyle\frac{1}{3!}\Bigg($
$\!\!\!1728~$
\hspace{-10pt}
\rule[-14pt]{0pt}{34pt}
\bpi(34,12)
\put(17,3){\circle{24}}
\put(6.6,9){\line(1,0){20.8}}
\put(6.6,9){\line(3,-5){10.4}}
\put(27.4,9){\line(-3,-5){10.4}}
\put(6.6,9){\circle*{4}}
\put(27.4,9){\circle*{4}}
\put(17,-9){\circle*{4}}
\epi
\hspace{10pt}
$\!\!\!3456~$
\hspace{-10pt}
\rule[-14pt]{0pt}{50pt}
\bpi(34,12)
\put(17,3){\circle{24}}
\put(17,3){\oval(24,8)}
\put(17,23){\circle{16}}
\put(5,3){\circle*{4}}
\put(29,3){\circle*{4}}
\put(17,15){\circle*{4}}
\epi
\hspace{10pt}
$\!\!\!2592~$
\hspace{-10pt}
\rule[-10pt]{0pt}{26pt}
\bpi(74,12)
\put(13,3){\circle{16}}
\put(29,3){\circle{16}}
\put(45,3){\circle{16}}
\put(61,3){\circle{16}}
\put(21,3){\circle*{4}}
\put(37,3){\circle*{4}}
\put(53,3){\circle*{4}}
\epi
\hspace{10pt}
$\!\!\!1728~$
\hspace{-10pt}
\rule[-18pt]{0pt}{50pt}
\bpi(53.7,12)
\put(26.85,3){\circle{16}}
\put(26.85,19){\circle{16}}
\put(13,-5){\circle{16}}
\put(40.7,-5){\circle{16}}
\put(26.85,11){\circle*{4}}
\put(19.95,-1){\circle*{4}}
\put(33.75,-1){\circle*{4}}
\epi
$\Bigg)$&9504
\\
\hline
$g^4$&
$\displaystyle\frac{1}{4!}\Bigg($
$\!\!\!62208~$
\hspace{-10pt}
\rule[-14pt]{0pt}{34pt}
\bpi(34,12)
\put(17,3){\circle{24}}
\put(8.5,-5.5){\line(1,0){17}}
\put(8.5,-5.5){\line(0,1){17}}
\put(8.5,11.5){\line(1,0){17}}
\put(25.5,-5.5){\line(0,1){17}}
\put(8.5,-5.5){\circle*{4}}
\put(8.5,11.5){\circle*{4}}
\put(25.5,-5.5){\circle*{4}}
\put(25.5,11.5){\circle*{4}}
\epi
\hspace{10pt}
$\!\!\!66296~$
\hspace{-10pt}
\rule[-22pt]{0pt}{50pt}
\bpi(42,12)
\put(21,-9){\circle{16}}
\put(21,15){\circle{16}}
\put(13,-9){\line(1,0){16}}
\put(13,15){\line(1,0){16}}
\put(13,3){\oval(16,24)[l]}
\put(29,3){\oval(16,24)[r]}
\put(13,-9){\circle*{4}}
\put(13,15){\circle*{4}}
\put(29,-9){\circle*{4}}
\put(29,15){\circle*{4}}
\epi
\hspace{10pt}
$\!\!\!248832~$
\hspace{-10pt}
\rule[-18pt]{0pt}{42pt}
\bpi(58,12)
\put(13,3){\circle{16}}
\put(45,3){\circle{16}}
\put(5,-5){\line(0,1){16}}
\put(25,-5){\oval(40,16)[b]}
\put(25,11){\oval(40,16)[t]}
\put(45,3){\oval(48,16)[l]}
\put(5,3){\circle*{4}}
\put(21,3){\circle*{4}}
\put(45,-5){\circle*{4}}
\put(45,11){\circle*{4}}
\epi
\hspace{10pt}
$\!\!\!497664~$
\hspace{-10pt}
\rule[-14pt]{0pt}{50pt}
\bpi(34,12)
\put(17,3){\circle{24}}
\put(17,23){\circle{16}}
\put(6.6,9){\line(1,0){20.8}}
\put(6.6,9){\line(3,-5){10.4}}
\put(27.4,9){\line(-3,-5){10.4}}
\put(6.6,9){\circle*{4}}
\put(27.4,9){\circle*{4}}
\put(17,-9){\circle*{4}}
\put(17,15){\circle*{4}}
\epi
\hspace{10pt}
$\!\!\!165888~$
\hspace{-10pt}
\rule[-14pt]{0pt}{47.3pt}
\bpi(46,12)
\put(23,3){\circle{24}}
\put(13,20.3){\circle{16}}
\put(33,20.3){\circle{16}}
\put(23,3){\oval(24,8)}
\put(11,3){\circle*{4}}
\put(35,3){\circle*{4}}
\put(17,13.4){\circle*{4}}
\put(29,13.4){\circle*{4}}
\epi
\hspace{10pt}
$\!\!\!248832~$
\hspace{-10pt}
\rule[-30pt]{0pt}{66pt}
\bpi(34,12)
\put(17,3){\circle{24}}
\put(17,-17){\circle{16}}
\put(17,23){\circle{16}}
\put(17,3){\oval(24,8)}
\put(5,3){\circle*{4}}
\put(17,-9){\circle*{4}}
\put(17,15){\circle*{4}}
\put(29,3){\circle*{4}}
\epi &
\\[-1.5cm]
&
$\!\!\!165888~$
\hspace{-10pt}
\rule[-14pt]{0pt}{66pt}
\bpi(34,12)
\put(17,3){\circle{24}}
\put(17,23){\circle{16}}
\put(17,39){\circle{16}}
\put(17,3){\oval(24,8)}
\put(5,3){\circle*{4}}
\put(17,15){\circle*{4}}
\put(17,31){\circle*{4}}
\put(29,3){\circle*{4}}
\epi
\hspace{10pt}
$\!\!\!124416~$
\hspace{-10pt}
\rule[-10pt]{0pt}{26pt}
\bpi(90,12)
\put(13,3){\circle{16}}
\put(29,3){\circle{16}}
\put(45,3){\circle{16}}
\put(61,3){\circle{16}}
\put(77,3){\circle{16}}
\put(21,3){\circle*{4}}
\put(37,3){\circle*{4}}
\put(53,3){\circle*{4}}
\put(69,3){\circle*{4}}
\epi
\hspace{10pt}
$\!\!\!248832~$
\hspace{-10pt}
\rule[-23.85pt]{0pt}{53.7pt}
\bpi(66,12)
\put(13,3){\circle{16}}
\put(29,3){\circle{16}}
\put(45,3){\circle{16}}
\put(53,-10.85){\circle{16}}
\put(53,16.85){\circle{16}}
\put(21,3){\circle*{4}}
\put(37,3){\circle*{4}}
\put(49,-3.9){\circle*{4}}
\put(49,9.9){\circle*{4}}
\epi
\hspace{10pt}
$\!\!\!62208~$
\hspace{-10pt}
\rule[-26pt]{0pt}{58pt}
\bpi(58,12)
\put(13,3){\circle{16}}
\put(29,-13){\circle{16}}
\put(29,3){\circle{16}}
\put(29,19){\circle{16}}
\put(45,3){\circle{16}}
\put(21,3){\circle*{4}}
\put(29,-5){\circle*{4}}
\put(29,11){\circle*{4}}
\put(37,3){\circle*{4}}
\epi   $ \Bigg)$&1880064
\\
\hline
\end{tabular}
\end{center}
\caption[Vacuum diagrams up to five loops and their
multiplicities.
]{Vacuum diagrams up to five loops and their
multiplicities.
}
\label{@vacdiagrkast}\end{figure}}


\end{document}